\documentclass{IEEEcsmag}

\usepackage[colorlinks,urlcolor=blue,linkcolor=blue,citecolor=blue]{hyperref}
\expandafter\def\expandafter\UrlBreaks\expandafter{\UrlBreaks\do\/\do\*\do\-\do\~\do\'\do\"\do\-}

\usepackage{upmath}
\usepackage{microtype}
\usepackage{amsmath,amssymb,amsfonts,bm}
\usepackage{graphicx}
\usepackage{booktabs}
\usepackage{tikz}
\usetikzlibrary{arrows.meta,positioning,calc,decorations.pathmorphing}
\usepackage{xcolor}
\usepackage[normalem]{ulem}
\newcommand{\added}[1]{{\color{black}#1}}

\jvol{XX}
\jnum{XX}
\paper{XX}
\jmonth{Month}
\jname{Computing in Science \& Engineering}
\jtitle{Controversies and Perspectives on AI/ML in Science \& Engineering}
\pubyear{2026}

\newcommand{\R}{\mathbb{R}}

\newcommand{\norm}[1]{\left\lVert #1\right\rVert}

\newcommand{\cM}{\mathcal{M}}
\newcommand{\cN}{\mathcal{N}}
\newcommand{\cG}{\mathcal{G}}

\newcommand{\cB}{\mathcal{B}}

\newcommand{\calN}{\mathcal{N}}
\newcommand{\calL}{\mathcal{L}}

\newcommand{\calG}{\mathcal{G}}

\newcommand{\calO}{\mathcal{O}}

\newcommand{\bfu}{\bm{u}}
\newcommand{\bfx}{\bm{x}}

\newcommand{\bftheta}{\bm{\theta}}

\begin{document}

\sptitle{Theme Article: Controversies and Perspectives in AI/ML}

\title{Predictivity and Utility of Neural Surrogates of Multiscale PDEs}

\author{Karthik Duraisamy}
\affil{University of Michigan}

\markboth{PREDICTIVITY AND UTILITY OF NEURAL PDE SURROGATES}{PREDICTIVITY AND UTILITY OF NEURAL PDE SURROGATES}

\begin{abstract}\looseness-1
Scientific machine learning is increasingly being spoken of as universal emulators for classical numerical solvers for multi-scale partial differential equations, but most apparent successes can be explained by  facts that also define their limits. Many successful benchmarks live on low-dimensional solution manifolds where any competent reduced model will interpolate well.  More fundamentally, neural surrogates systematically under-resolve high-frequency content due to spectral bias, and coarse-graining compounds this problem through irreversible information loss. In many multi-scale problems, no architecture or training procedure can fully recover what the coarse representation discards. Two simple examples are used to characterize spectral bias, coarse-graining and error accumulation. We discuss why medium-range weather prediction on reanalysis data sits in a favorable sweet spot and why this will not generalize to genuinely chaotic multi-scale scenarios. We identify domains where neural surrogates offer genuine value, propose further research on neural-classical hybrids, and call for better reporting standards.
\end{abstract}

\maketitle

\chapteri{S}cientific machine learning (SciML) has produced a fast-growing zoo of neural surrogates for multiscale partial differential equations (PDEs), often accompanied with maximalist claims such as ``replace the solver'' or ``1000$\times$ speedups.'' Many  successes reflect one of two situations:

The first situation is \emph{interpolation on a benign manifold}. When the parameter-to-solution map is smooth and the solution set has low intrinsic dimension: rapidly decaying Kolmogorov n-width \cite{Pinkus1985}, any competent approximation method will succeed. Polynomial surrogates, projection-based Reduced Order Models (ROMs), or neural networks all achieve similar accuracy on such problems.  The key diagnostic is whether classical reduced-order methods with comparable offline cost achieve similar accuracy \cite{McGreivyHakim2024}; if so, the benchmark tests interpolation, not the capacity to handle multiscale physics.

The second situation is \emph{confusing attractor sampling with prediction}. Showing that a surrogate reproduces mean profiles or energy spectra on a chaotic attractor demonstrates that it has learned to sample the invariant measure \cite{Linot2023}. This is useful for estimating long-time averages but categorically different from forecasting: the model may have no skill at predicting which region of the attractor the system will visit next. The distinction between learning dynamics and learning statistics is well-understood in dynamical systems but often glossed over in SciML benchmarks.

An important issue cuts across both of the aforementioned situations: neural networks impose an implicit \emph{low-pass prior} on what they learn, {a phenomenon known as \emph{spectral bias}}. Rahaman et al.\ \cite{Rahaman2019}  showed that networks trained by gradient descent fit low-frequency components first,  a phenomenon now understood through the Neural Tangent Kernel (NTK) \cite{jacot2018neural}. {In essence, the kernel eigenvalues associated with high-frequency Fourier modes decay polynomially, so that gradient-based training resolves low-frequency structure orders of magnitude faster than fine-scale content.} Wider networks shift the spectrum but generally preserve the decay rate; more data cannot create gradient signal for modes where the eigenvalue is effectively zero. Architectural choices make this explicit: Fourier neural operators truncate to $k_{\max}$ modes by design \cite{Li2021FNO}, and even ``super-resolution'' variants cannot recover information never represented.

For PDEs, spectral bias is not an aesthetic defect as in image processing: high-$k$ content encodes boundary layers, sharp fronts, and  fluctuations that control derivative-dependent quantities of interest: e.g. fluxes, stresses, dissipation rates. A surrogate can achieve excellent $L^2$ error while systematically misrepresenting the physics that matters. This explains persistent reports of PINNs failing on ``slightly harder'' problems and neural operators degrading under autoregressive rollout \cite{Lippe2024}.

Coarse-graining introduces a deeper problem: \emph{irreversible information loss}. A network trained with mean-squared error converges to the conditional expectation $\hat{u}(u_c) = \mathbb{E}[u \mid u_c]$ \cite{wang2024coarsegraining}, i.e. the average over all fine states consistent with the coarse input. This is simply the \emph{optimal} solution to Mean Squared Error (/MSE) minimization . The resulting prediction is over-smoothed by construction, and no architecture or training procedure can recover information the coarse representation discards. 

{We note that many of these challenges (e.g. closure modeling, projection-induced information loss, and instability under long-horizon rollout) are well established in the classical ROM literature, and comments above apply equally to any coarse representation, neural or otherwise. The distinction is that classical intrusive ROMs retain explicit access to the governing equations {\em in discrete form}, whereas neural surrogates trade this structural access for flexibility and non-intrusiveness; the hybrid strategies discussed later in this paper can be viewed as attempts to recover some of these structural guarantees by periodically re-coupling the surrogate with a physics-based solver.}

These observations reframe how we should evaluate neural PDE surrogates.  Related issues have been raised by others as well (e.g.~\cite{McGreivyHakim2024,Brunton2024}). In this work, we provide some partial answers and focus on : \emph{What is being predicted: trajectories, statistics, or merely interpolated parameter responses? On what horizon, measured in relevant physical time scales? And do the metrics probe the scales that control the quantities of interest, or do they hide spectral deficiencies behind smooth headline numbers?}


\added{This paper illustrates with simple examples, some interacting issues that limit predictive capabilities of neural surrogates for multiscale PDEs:  \emph{spectral bias}, \emph{coarse-graining information loss}, and \emph{autoregressive error accumulation}}. We also identify problem classes where neural surrogates offer genuine value and discuss hybrid neural--classical strategies as a path forward. We distinguish two evaluation criteria: \emph{predictivity} (accuracy on physically relevant quantities of interest under extrapolation) and \emph{utility} (end-to-end cost advantage over classical methods), arguing that conflating the two has led to overoptimistic assessments.

\noindent {\bf Predictivity and Utility} We try to separate two questions that are often conflated.
\emph{Predictivity} asks whether a surrogate remains accurate under the extrapolations that matter for the scientific task: longer horizons, distribution shift, and/or resolution refinement, as measured on physically relevant Quantities of Interest (QoI), often derivative- or phase-sensitive).
\emph{Utility} asks whether the surrogate reduces end-to-end \emph{time-to-QoI} relative to optimized solvers and classical reduced models once training cost, inference cost, and amortization over the intended query count are included.
A method can be useful without being truly predictive (e.g., accurate long-time statistics with  trajectory errors), and it can be predictive without being useful (e.g., accurate but not competitive end-to-end from a cost consideration standpoint).  Also, these distinctions vary from task to task. In design optimization, end-to-end cost is extremely important. In short term weather prediction, only the inferencing cost is relevant. This paper exposes some of these issues in a simple setting (codes available at {\em  https://github.com/CaslabUM/NeuralSurrogates}), and provides some perspectives on how to move forward.

	\section{A frequency-response lens on PDE surrogates}
	
{The interplay between spectral bias and PDE structure is best understood by examining how three frequency-dependent quantities interact. This section introduces a unifying lens by connecting the PDE's own spectral content, the operator's frequency response, and the learning dynamics of the network. This lens will be used  to diagnose when and why neural surrogates succeed or fail.}

For PDE surrogates, predictive ability becomes explanatory only when combined with the PDE's own frequency response and with the metric/QoI being reported. Three ``spectra'' interact: {\em (i) Solution spectrum} $|\hat u(k)|$: what scales the PDE actually contains (and which scales are energetically significant); 
		 {\em (ii) Operator spectra} $\sigma_{\calL}(k)$: how the PDE maps forcing/initial data to solution components in Fourier space; 
	and	 {\em (iii) ML spectra} $\lambda_k$ (NTK eigenvalues / architectural response): how quickly training reduces error in each mode.

	{For a linear constant-coefficient PDE $\calL u = f$ on a periodic domain, the Fourier transform  gives $\sigma_{\calL}(k)\,\hat u(k)=\hat f(k)$, where $k$ is wavenumber, and hence}
	\begin{equation}
		\hat u(k)=\frac{\hat f(k)}{\sigma_{\calL}(k)}.
		\label{eq:pde_symbol}
	\end{equation}
	{If $|\sigma_{\calL}(k)|$ grows with $|k|$ (elliptic/parabolic smoothing), then high-$k$ content in $u$ is naturally small, and a low-pass learner can still be useful for low-order QoIs. If $|\sigma_{\calL}(k)|$ does \emph{not} grow (transport/waves/Helmholtz at high frequency) or if nonlinear dynamics continuously regenerate fine scales (turbulence), then under-resolving high $|k|$ directly biases quantities controlled by gradients, dissipation, or phase.}
	
	{The loss/metric introduces another frequency weighting. For a linear PDE residual loss on a periodic domain,}
	\begin{equation}
		\|\calL u_\theta - f\|_{L^2}^2
		= \sum_k |\sigma_{\calL}(k)|^2\,|\hat u_\theta(k)-\hat u(k)|^2,
		\label{eq:residual_weighting}
	\end{equation}
	whereas derivative-sensitive QoIs correspond to scaled weights.
	Thus ``small MSE'' or ``small residual'' can coincide with large errors in the wavenumber bands that dominate dissipation or stability. This is the precise bridge from the NTK analysis below (frequency-dependent convergence rates) to the PDE deficits discussed later (multiscale stiffness and long-horizon failure).

\section{Spectral Bias: The Neural Tangent Kernel Perspective}

{Having introduced the frequency-response framework, we now formalize the source of spectral bias. This section develops the theory and demonstrates it on a controlled numerical example.  The NTK framework \cite{jacot2018neural} provides a rigorous foundation for understanding spectral bias: the kernel's eigenvalues decay polynomially with frequency, so that high-wavenumber modes are learned orders of magnitude more slowly than low-wavenumber modes.  The implication for predictivity is direct: surrogates may appear accurate on $L^2$ metrics while systematically under-resolving the scales that govern derivatives, dissipation, and phase.}

Consider a neural network $f: \R^d \times \R^p \to \R$ that maps input $\bfx \in \R^d$ to output $f(\bfx; \bftheta)$, parameterized by weights $\bftheta \in \R^p$. We train on data $\{(\bfx_i, y_i)\}_{i=1}^n$ by minimizing the mean squared error:
\begin{equation}
	\mathcal{L}(\bftheta) = \frac{1}{2n} \sum_{i=1}^n \left( f(\bfx_i; \bftheta) - y_i \right)^2.
\end{equation}

The NTK is the positive semi-definite kernel $K: \R^d \times \R^d \to \R$ defined by:
\begin{equation}
	K(\bfx, \bfx') = \left\langle \nabla_{\bftheta} f(\bfx; \bftheta), \nabla_{\bftheta} f(\bfx'; \bftheta) \right\rangle = \sum_{i=1}^p \frac{\partial f(\bfx; \bftheta)}{\partial \theta_i} \frac{\partial f(\bfx'; \bftheta)}{\partial \theta_i}.
\end{equation}

The key insight~\cite{jacot2018neural} is that, for networks trained with gradient descent, the NTK eigenvalues $\lambda_k$ associated with Fourier mode $k$ decay polynomially:
\begin{equation}
	\lambda_k \sim k^{-\alpha}, \quad \alpha \approx 2 \text{ for smooth activations.}
\end{equation}

{If the NTK eigenvalues scale like $\lambda_k\sim k^{-\alpha}$, then reaching a fixed relative accuracy at frequency $k$ requires training time $t_k=\calO(k^\alpha)$. Equivalently, for a fixed training budget $t$ there is an implicit ``resolved cutoff''}
\begin{equation}
	k_{\max}(t,\epsilon)\ \approx\ \Big(\eta t/\ln(1/\epsilon)\Big)^{1/\alpha},
	\label{eq:km_cutoff}
\end{equation}
{beyond which modes remain largely unlearned~\cite{ronen2019convergence}. For the empirically observed $\alpha\approx 2$, $k_{\max}$ grows only like $\sqrt{t}$: each extra decade of resolvable wavenumbers costs roughly $100\times$ more iterations.}



\begin{figure*}
	\centering
	\includegraphics[width=\textwidth]{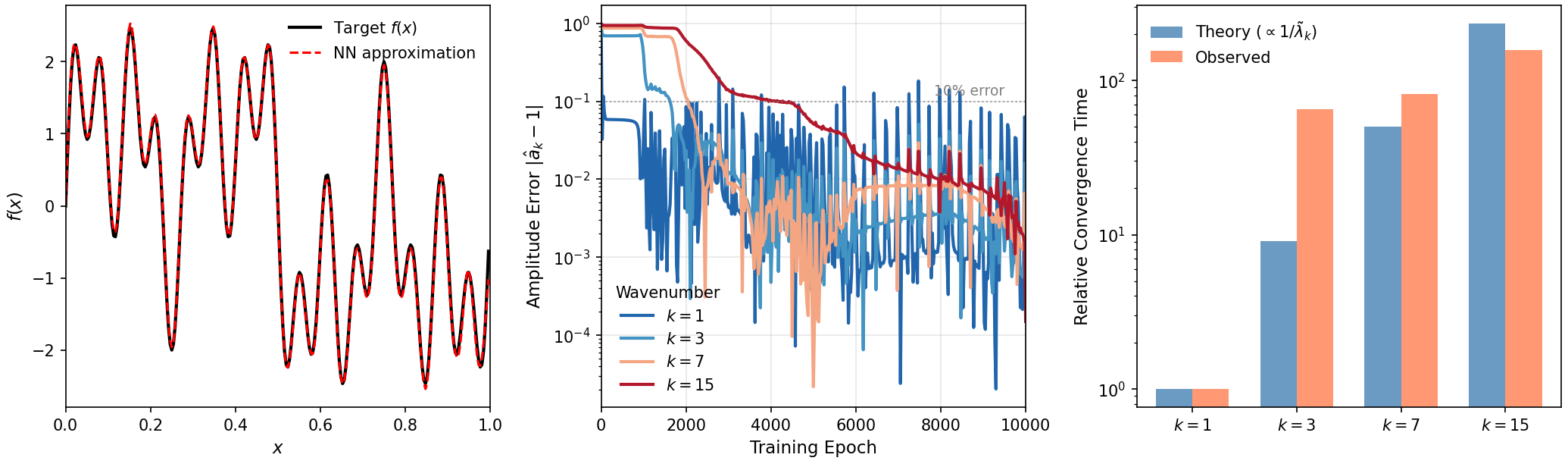}
	\caption{Learning dynamics with equal-amplitude target. (a) Target function (all amplitudes = 1.0) and final NN approximation. (b) Per-wavenumber convergence: despite equal target amplitudes, high wavenumbers converge much slower. (c) Comparison of theoretical predictions ($\propto 1/\lambda_k$) with observed convergence times.}
	\label{fig:learning_dynamics}
\end{figure*}

To illustrate this concept, we construct a target with \textbf{equal amplitudes} across all frequency components:
\begin{equation}
	f(x) = \sin(2\pi x) + \sin(6\pi x) + \sin(14\pi x) + \sin(30\pi x).
\end{equation}
Each Fourier component ($k = 1, 3, 7, 15$) has amplitude exactly 1.0.
We  compute the NTK for a 4-layer neural network with 128 hidden units and $\tanh$ activations. For a grid of $N = 256$ points on $[0,1]$, we approximate Fourier eigenvalues $\{\tilde{\lambda}_k\}$ via Rayleigh quotients.
\begin{figure}
	\centering
	\includegraphics[width=0.45\textwidth]{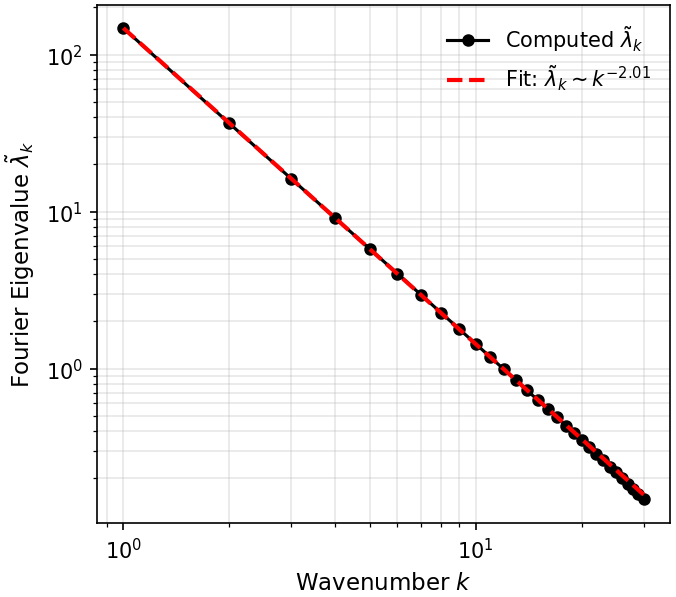}
	\caption{Neural Tangent Kernel eigenvalue analysis.  Fourier eigenvalues $\tilde{\lambda}_k$ versus wavenumber}
	\label{fig:ntk_eigenvalues}
\end{figure}
Figure~\ref{fig:ntk_eigenvalues} shows the key result: the Fourier eigenvalues decay as $\tilde{\lambda}_k \sim k^{-2.01}$, essentially exactly $k^{-2}$ as predicted by theory for smooth activations. 
The agreement between computed ratios and the $k^{-2}$ prediction is striking, with relative errors below 2\%.   Figure~\ref{fig:learning_dynamics} shows the learning dynamics.

Figure~\ref{fig:learning_dynamics} compares the observed convergence times with the theoretical prediction based on NTK eigenvalues. The theory predicts convergence time $\propto 1/\lambda_k$. The agreement is reasonable, confirming that the NTK framework captures the essential mechanism. Despite having identical amplitude in the target, the $k=15$ mode requires $158\times$ more training iterations than the $k=1$ mode.

For operator learning, spectral bias compounds through input encoding, operator approximation, and output decoding. The Fourier Neural Operator's explicit truncation to $k_{\max}$ modes makes this ceiling explicit. Proponents argue that spectral bias can be overcome with larger models and more data. This is partially true but fundamentally limited:

\noindent \textbf{Width scaling}: Wider networks do not change the NTK eigenvalue \emph{decay rate}~\cite{Bietti19}: they shift the spectrum but preserve $\lambda_k \propto k^{-\alpha}$.

\noindent  \textbf{Data scaling}: More training data helps accuracy at resolved scales but cannot create signal for modes where the NTK eigenvalue is effectively zero~\cite{ronen2019convergence}.

\noindent \textbf{The inertial range barrier}: For turbulence with $k_{\max}/k_{\min} \sim 10^4$, the implied $\sim 10^8\times$ separation in learning time across scales means no practical training budget achieves uniform accuracy.

{A substantial body of work has sought to mitigate spectral bias through input encodings such as random Fourier features~\cite{tancik2020fourier} and sinusoidal representations, adaptive activation functions, multi-scale architectures, and frequency-progressive training curricula, all of which can flatten the effective NTK spectrum and accelerate high-frequency convergence on problems where the relevant wavenumber range is known \emph{a priori}. These techniques can help in fixed-spectrum settings (e.g., static reconstruction, single-PDE benchmarks). For broadband, dynamically evolving spectra, no fixed encoding or curriculum can anticipate the frequencies that will matter at future times, and the cost of uniformly resolving all scales still grows polynomially with the spectral bandwidth. Thus, mitigations {\em can } help, but they do not eliminate the  competition between finite training budgets and  resolution demands of genuinely multi-scale PDEs.}

\section{What is being optimized?}

{The previous section showed that gradient-based training inherently favors low frequencies.}
A common intuition is that enforcing the PDE loss $\mathcal{L}u_\theta \approx f$ should ``force'' high-frequency accuracy. In practice, the situation is subtler because the training objective induces \emph{frequency-dependent conditioning} that can amplify the mismatch between what the PDE demands and what the network learns easily. For linear $\mathcal{L}$, the residual loss $\|\mathcal{L}u_\theta-f\|_2^2$ weights Fourier modes as
\[
\|\mathcal{L}u_\theta-f\|_2^2
\;=\;
\int |\sigma_{\mathcal{L}}(k)|^2\,|\widehat{u_\theta}(k)-\widehat{u}(k)|^2\,dk.
\]
Thus, the PDE operator reweights the solution error by $|\sigma_{\mathcal{L}}(k)|^2$. This can be beneficial (some operators penalize high-$k$ error more), but it can also make the optimization \emph{stiff} across scales: if certain frequencies are heavily weighted by the physics, but those same frequencies correspond to directions that gradient-based training learns slowly in the model class, training can plateau with a deceptively small residual and a large band-limited solution error. 

A physics informed neural network~\cite{Raissi2019} (PINN) posits $u_\theta(x,t)$ and minimizes
\begin{equation}
	\mathcal{L}(\theta)
	=
	\norm{ \cN_\mu(u_\theta)}^2_{\Omega_c}
	+\norm{\cB_\mu(u_\theta)}^2_{\partial\Omega_c}
	+\norm{u_\theta-y}^2_{\Gamma},
	\label{eq:pinn_loss}
\end{equation}
where $\Omega_c$ are collocation points, and $\Gamma$ denotes sensor/data points.
For multiscale PDEs, the residual contains derivatives that amplify high frequencies.
If $u(x)=\sum_{k\in\mathbb{Z}^d} \hat u_k e^{ik\cdot x}$ on a periodic domain, then
\begin{equation}
	\norm{\nabla(u-\widehat{u})}_{L^2}^2 = \sum_k \norm{k}^2\,|\hat u_k-\widehat{u}_k|^2,
	\label{eq:H1_weighting}
\end{equation}
A growing literature documents that such stiffness and multiscale structure can cause naive PINN training to stall or converge to qualitatively wrong solutions, even on problems that are ``only slightly'' more complex than canonical toy examples so small errors at large $|k|$ are disproportionately important for derivative-based physics.
A ``low-frequency-correct'' PINN can have acceptable $L^2$ error yet high  errors in quantities of interest. This perspective also motivates frequency-aware preconditioning (e.g., loss reweighting or multiresolution curricula) as principled responses to spectral mismatch.

Operator learning aims to approximate a map
\begin{equation}
	\cG:\ a \mapsto u,
	\label{eq:operator_map}
\end{equation}
where $a$ could be a coefficient field, forcing, geometry encoding, etc.
Neural operators~\cite{Li2021FNO} discretize $a$ and output $u$ on a grid.
Most current successes can be interpreted as learning an efficient representation of $\cG$ on a distribution where $\cM$ is low-dimensional and smooth.
Crucially, for chaotic systems, the map $u_0\mapsto u(T)$ becomes heavily non-Lipschitz as $T$ grows.
Thus operator learning for long rollouts is not likely to result in good trajectory predictions, and even statistics may not be well-predicted if the impact of high wavenumber dynamics is significant.

\section{When are Errors Manageable: The Sweet Spot Problems}

{The preceding sections may give the impression that spectral bias renders neural surrogates universally unreliable. That is not the case. There exist important problem classes where spectral bias is either benign or actively aligned with the physics. Recognizing these ``sweet spots'' is essential for deploying surrogates where they offer genuine utility. This section identifies three such classes and then dissects why medium-range weather prediction on reanalysis data sits in a particularly favorable regime.}

\noindent \textbf{Steady-State/Elliptic Problems.}
For $-\Delta u = f$, solutions gain two derivatives of regularity: $|\hat{u}_k| \lesssim |k|^{-(s+2+d/2)}$ vs. $|\hat{f}_k| \lesssim |k|^{-(s+d/2)}$.
The spectral bias \emph{aligns} with natural spectral decay---high-frequency content is already small.

\noindent \textbf{Diffusion-Dominated Problems.}
For $u_t = \nu \Delta u$, high-frequency modes decay as $\hat{u}_k(t) = \hat{u}_k(0) e^{-\nu |k|^2 t}$.
Physical damping complements spectral bias; the ``true'' solution becomes progressively smoother.

\noindent \textbf{Short-Time Predictions with Smooth Data.}
Nonlinear energy transfer to high wavenumbers takes time. For smooth initial data, solutions remain smooth on $[0, T^*]$ before fine-scale structure develops.

\subsection*{The ERA5: Anatomy of a Sweet Spot}

Given the  complexity of the underlying physics, the success of neural weather prediction on ERA5 reanalysis~\cite{Hersbach2020ERA5} deserves careful dissection: a) At 25 km resolution, the atmosphere is a smooth, filtered representation~\cite{Bolgiani2022ERA5} with effective low $Re \sim 10^{3}$--$10^{4}$ (matched to the resolution), not the true $Re \sim 10^{12}$;
b) 1-10 day forecasts correspond to $\mathcal{O}(1)$ planetary-wave turnovers, evaluated \emph{before} high-frequency errors propagate upward via baroclinic error growth; c) RMSE weights large-scale patterns; a forecast missing all mesoscale structure scores excellently. Also note that ERA5 is itself spectrally biased (data assimilation trusts large scales). Training and evaluating on ERA5 creates a favorable match.

\textbf{The Lyapunov Time Perspective.} It is helpful to translate the usual ``lead time in days'' into nondimensional units that are
physically meaningful for atmospheric predictability~\cite{Vannitsem2017Predictability}.
For atmospheric models, leading Lyapunov exponents are $\lambda_1 \sim 0.2$--$0.4\,\mathrm{day}^{-1}$, giving error-doubling times $T_2 \sim 2$--$3$ days \cite{snyder2003lyapunov}.
A 10-day forecast spans $\sim 3$--$5$ doubling times beyond which \emph{any} small-scale error saturates.
{\bf The celebrated metrics evaluate exactly the window where low-frequency skill dominates and before spectral bias consequences manifest.} Another unusual aspect of ERA5 is that a large amount of high quality structured,  data is available for the very system/geometry that is the target.  This is unusual in almost any problem in science or engineering.

The author is careful not to dismiss weather forecasting as an easy or solved problem: This is indeed, an example of an extremely successful application of neural surrogates with significant real-world impact. In the present context, ERA5-based weather prediction scores high on both utility (orders-of-magnitude inference speedup over numerical weather prediction) and predictivity (skillful forecasts on the metrics and horizons that matter operationally). \added{Understanding \emph{why} this sweet spot exists, e.g. filtered data, favorable Lyapunov horizons, large-scale-dominated metrics, is essential to assess whether similar success will transfer to other domains.} Further work is required to achieve super-resolved forecasts and in predicting origins and trajectory of extreme weather events, both of which are active research areas.

\section{Challenges in Chaotic Multi-Scale Dynamics}

{We now turn from the favorable cases to the genuinely hard ones. } As a prototypical multi-scale problem, turbulent flows exhibit fundamentally different spectral characteristics.
The energy spectrum follows Kolmogorov scaling $E(k) \sim \varepsilon^{2/3} k^{-5/3}$ in the inertial range, with scale separation $k_\eta/k_f \sim Re^{3/4}$. In many engineering and science problems, this is $\sim 10^6$.

Unlike diffusion-dominated flows, turbulence \textbf{continuously regenerates} high-frequency content through nonlinear interactions.
Any mechanism that removes energy at scale $k$ disrupts the cascade equilibrium; the flow generates new fine scales that the biased approximation again fails to capture.

The deepest issue is not approximation error but \textbf{irreversible information loss}. Coarse graining defines an observation $u_c = Pu$ where $P$ projects onto resolved modes.
If $\ker(P) \neq \{0\}$, multiple fine states share the same coarse representation:
\begin{equation}
Pu_1 = Pu_2 \quad \text{but} \quad u_1 \neq u_2.
\label{eq:nonidentifiable}
\end{equation}

A deterministic model trained with MSE converges to the \textbf{conditional mean}:
\begin{equation}
\hat{u}(u_c) = \mathbb{E}[u \mid u_c],
\end{equation}
which is the ``over-smoothed'' field that averages over unresolved high-frequency variability.
This is not a training failure---it is the \emph{optimal} MSE solution given the information available.

The Mori-Zwanzig formalism~\cite{Chorin2000} makes this precise: the exact reduced dynamics contains memory kernels and stochastic forcing that encode the influence of unresolved scales.
A deterministic neural surrogate trained on coarse data is \emph{fundamentally unable} to recover this structure as it learns the wrong physics, not just an imprecise approximation.

Wang et al.\ \cite{wang2024coarsegraining} propose bypassing closure models entirely using physics-informed neural operators (PINO) that operate in function space at fine resolution. Their approach trains PINO first on coarse-grid solver data, then fine-tunes with a small amount of fully-resolved simulation data and physics-based losses on the fine grid. Because neural operators are discretization-free, they are not constrained to the coarse grid that makes closure modeling ill-posed. The insight is data efficiency: learning-based closure models require enough data that the dataset itself suffices to estimate statistics directly. But predictivity in different problems remains an issue.

\subsection{The Accumulation Problem}

{Error accumulation is the primary mechanism by which spectral bias degrades predictivity over time.}
Consider a neural time-stepper $\bfu^{n+1} = \calN_\theta(\bfu^n)$ with per-step error $\bm{\epsilon}^{n+1} = \calN_\theta(\bfu^n) - \calN^*(\bfu^n)$.
Due to spectral bias, $\hat{\bm{\epsilon}}_k$ is systematically larger for high $|k|$.
In chaotic systems with positive Lyapunov exponents $\lambda_1 > 0$, errors grow as $\|\bm{\epsilon}_{\text{total}}\| \sim e^{\lambda_1 N \Delta t} \|\bm{\epsilon}\|$. Therefore,
for any PDE with positive Lyapunov exponent and broadband spectrum, one can expect  a finite $T^*$ such that trajectory error exceeds any threshold, regardless of training.

\begin{figure*}[t]
	\centering
	\includegraphics[width=0.4\linewidth]{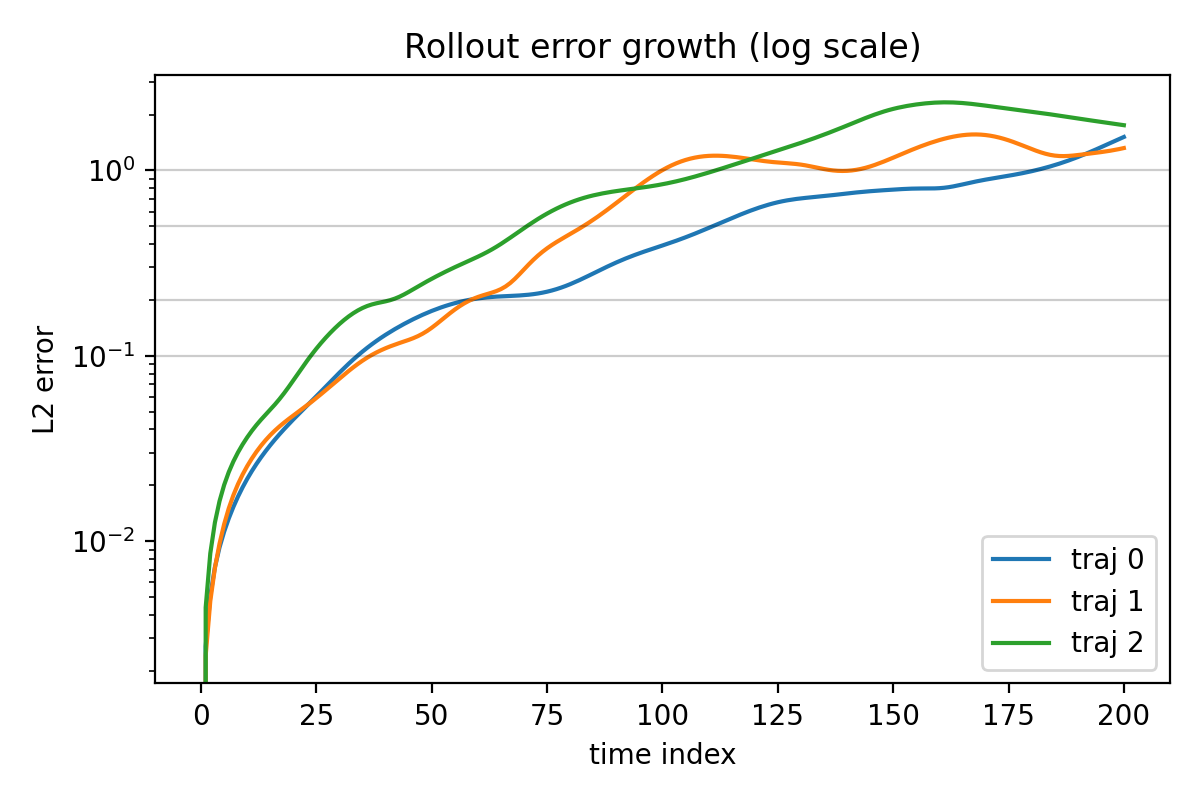}\\
	\includegraphics[width=0.3\linewidth]{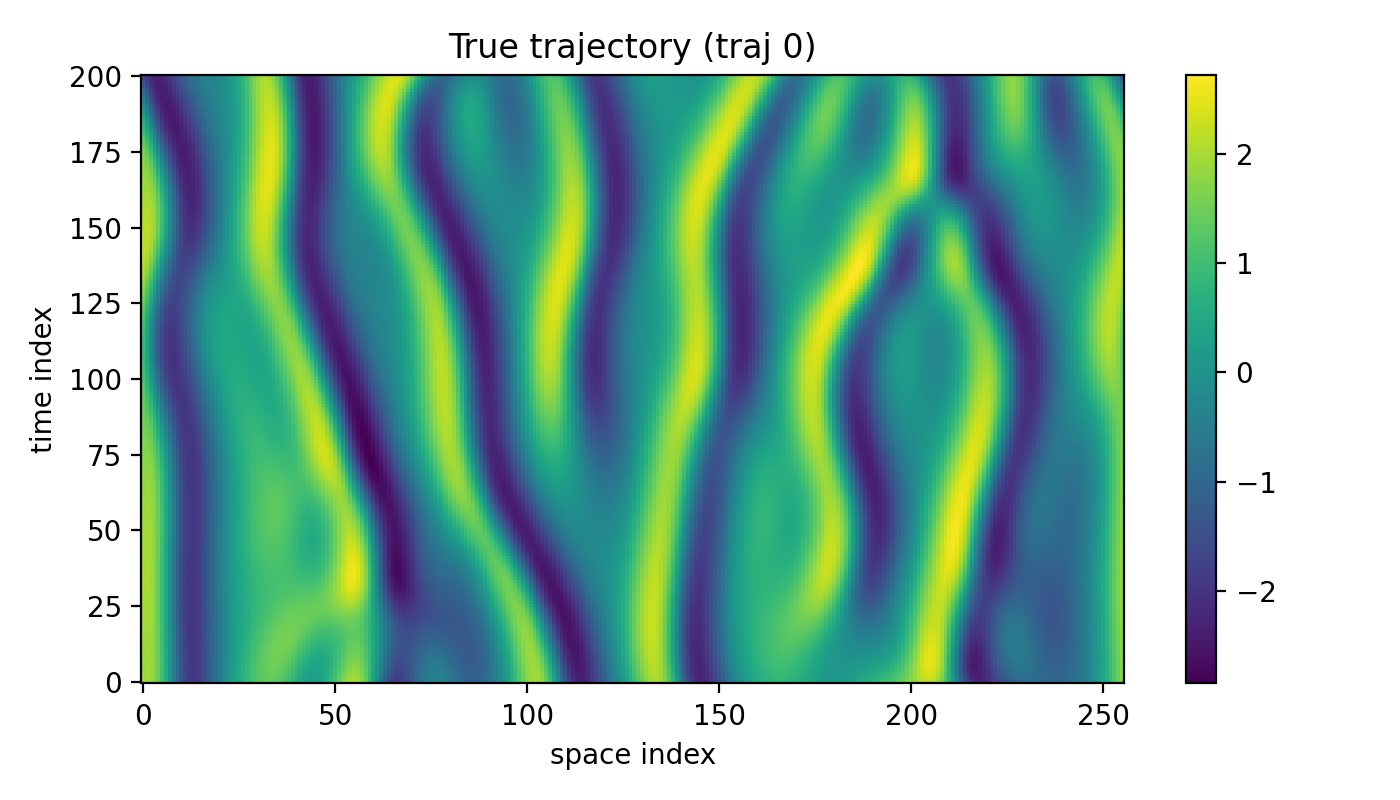}\hfill
	\includegraphics[width=0.3\linewidth]{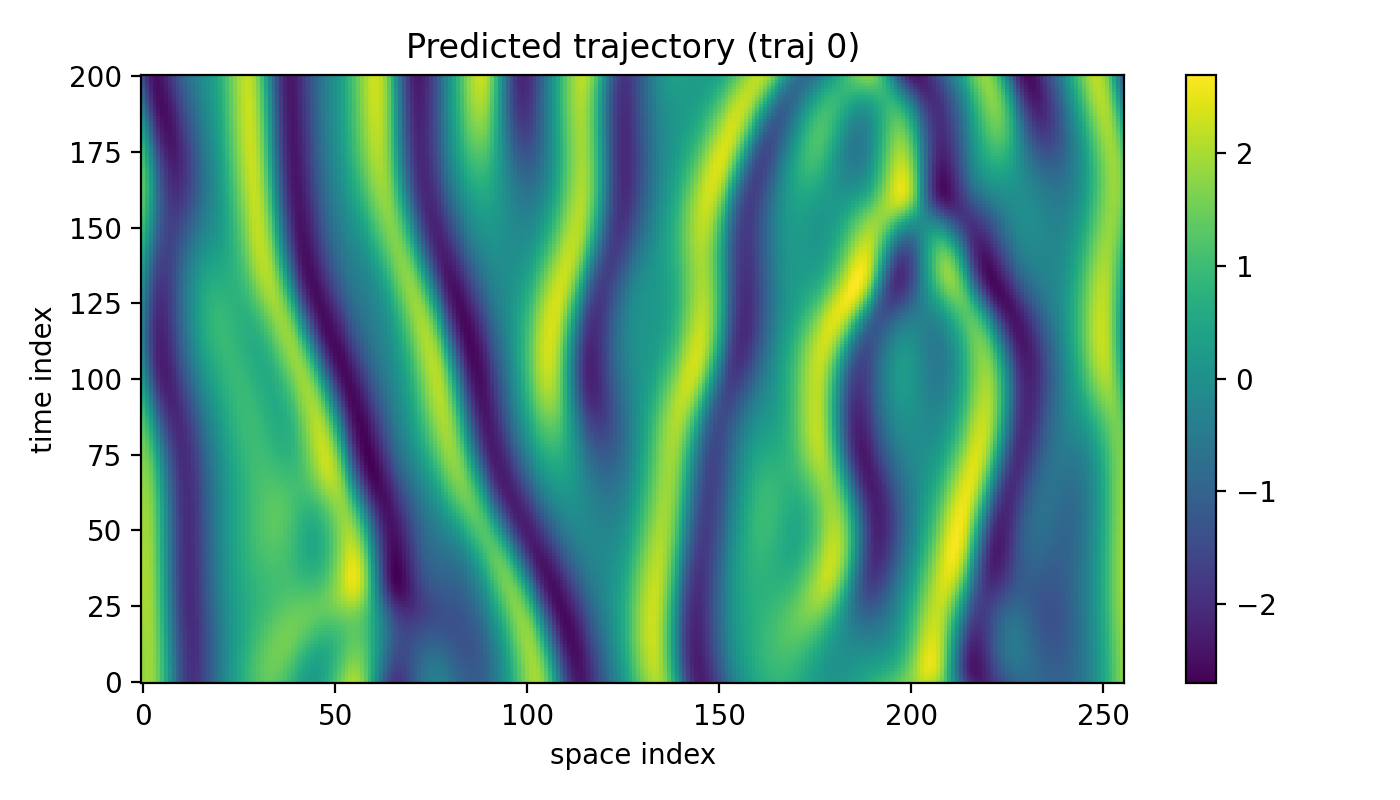}\hfill
	\includegraphics[width=0.3\linewidth]{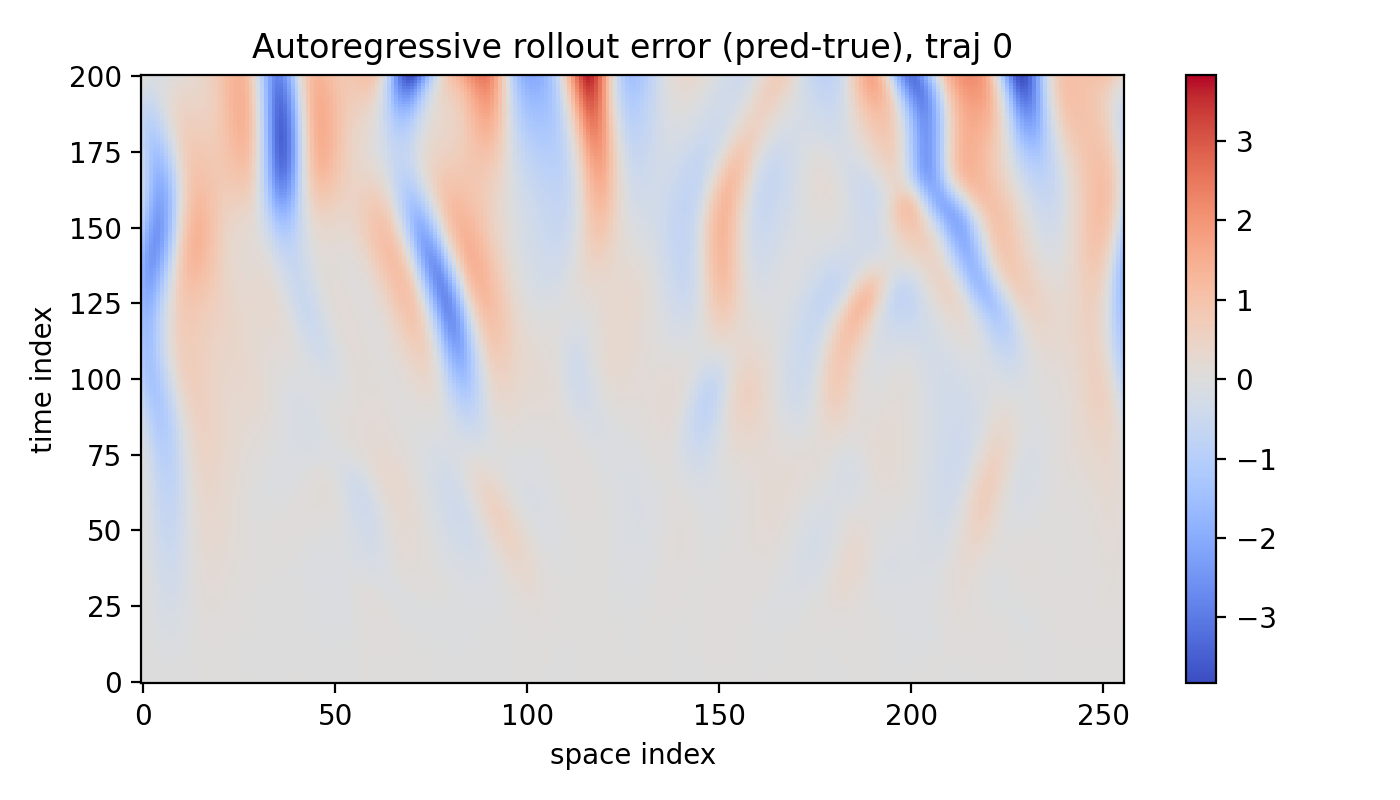}\\
	\includegraphics[width=0.4\linewidth]{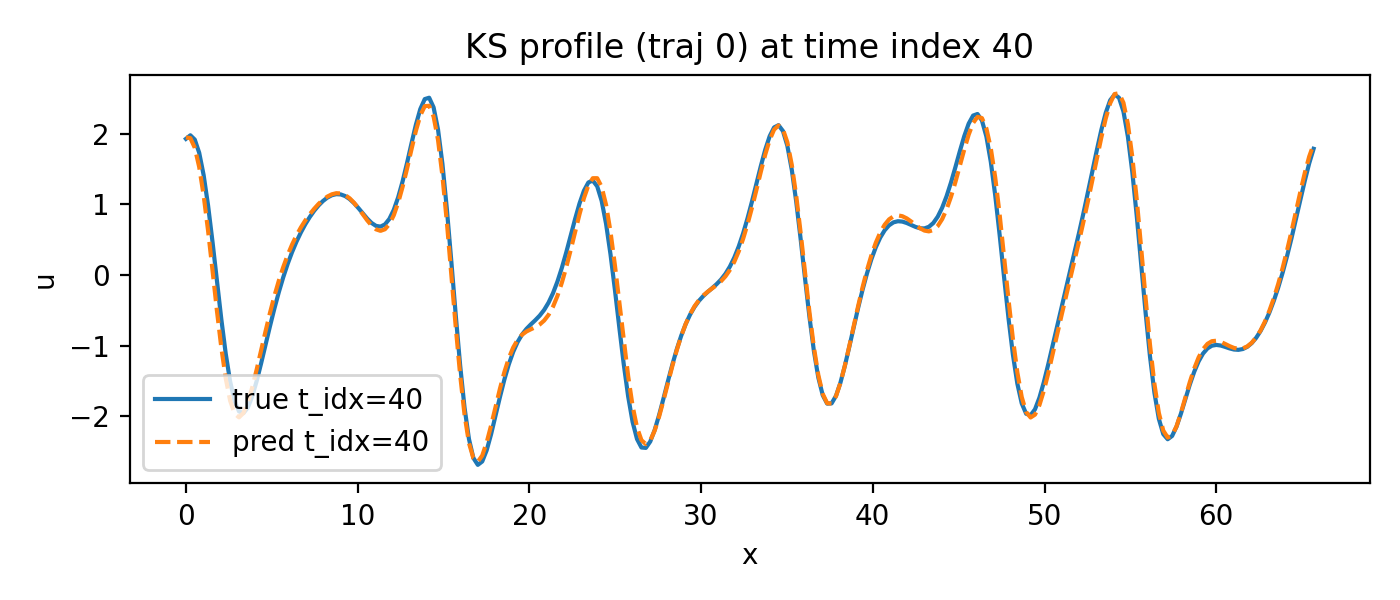}\hfill
	\includegraphics[width=0.4\linewidth]{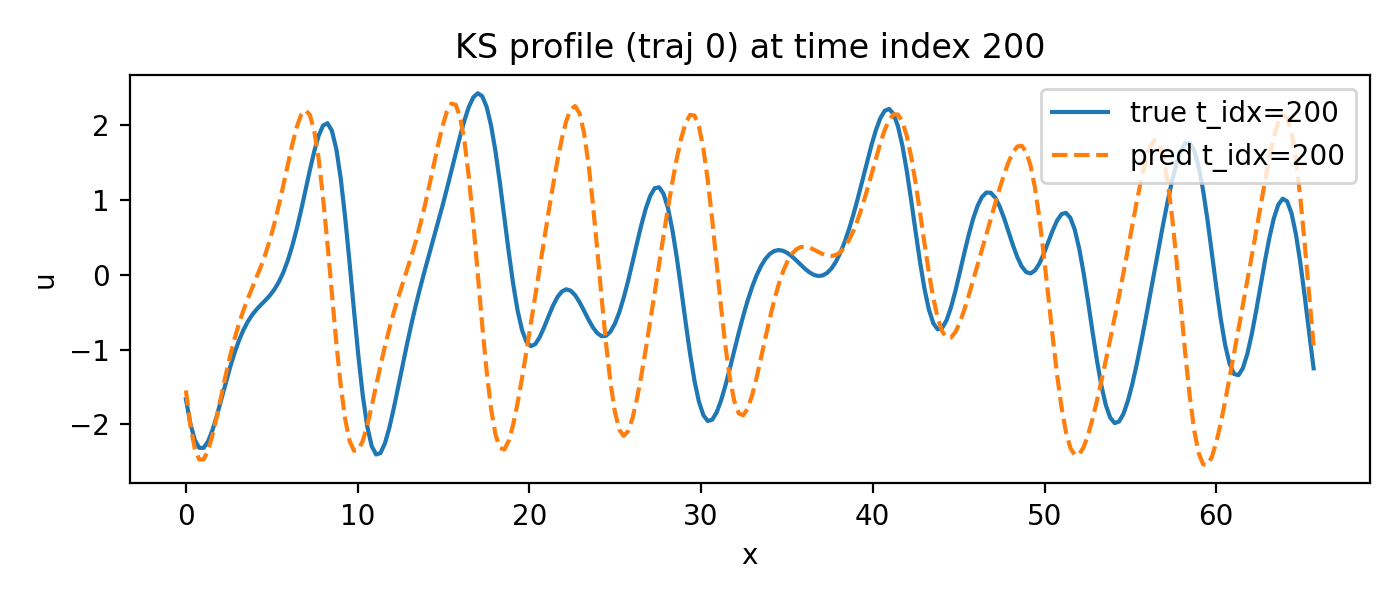}\\
	\includegraphics[width=0.4\linewidth]{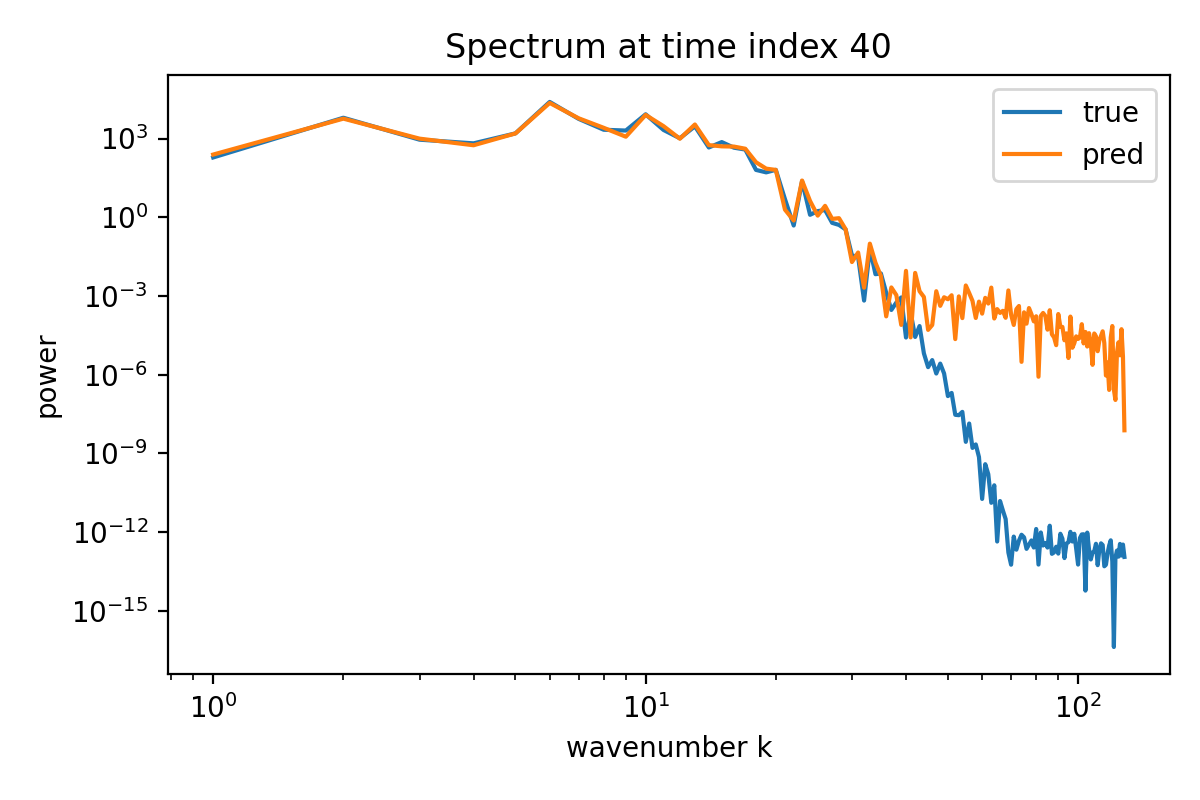}\hfill
	\includegraphics[width=0.4\linewidth]{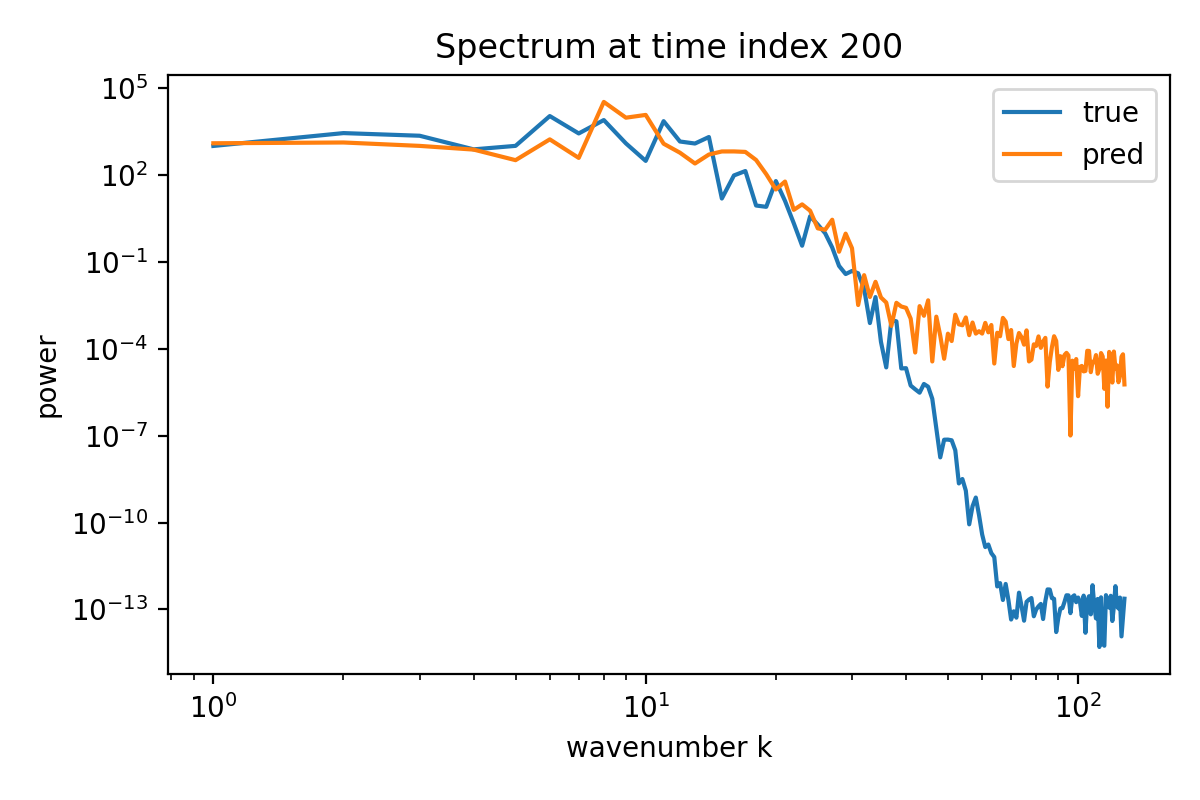}
	\caption{Rollout mean squred error $\mathrm{L2}(t)$ for three test trajectories, Rollout heatmaps, profile and spectra (traj 0).}
	\label{fig:ks}
\end{figure*}

As an illustration, we show a 1D Kuramoto--Sivashinsky (KS) equation using a Fourier neural operator (FNO) surrogate. Dataset and architectural details are the same as the baseline case in Lippe et al.~\cite{Lippe2024}.  We train a one-step surrogate $\hat{u}_{t+1}=\calG_{\bftheta}(u_t)$ on KS trajectories, then evaluate it in \emph{autoregressive} rollouts.
Even when one-step error is small, rollout error  grows quickly in chaotic or multiscale regimes as seen in Figure~\ref{fig:ks}. This case study highlights an important caveat:
even if the network fits low/mid $k$ extremely well, the high-$k$ modes may never be driven to the physically correct level if (i) they are weakly constrained by data (true spectrum already below noise) and/or (ii) the effective kernel has tiny eigenvalues in those directions.

For operator learning, a second complication is rollout:
inputs shift from true states $u_t$ (training distribution) to predicted states $\hat{u}_t$.
Thus, even if NTK arguments explain which modes are learned quickly in the supervised one-step setting, they do not automatically guarantee stability under distribution shift.
The observed frequency-selective mismatch (large $\Delta(k,t)$ at high $k$) provides a concrete mechanism for rollout degradation.

 Let $E_B(t)=\sum_{k\in B}|\widehat u(k,t)|^2$ denote the total spatial Fourier energy in a band $B$.
We define a band-energy relative error
\begin{equation}
	\varepsilon_B(t)=\frac{|E_{B,\mathrm{pred}}(t)-E_{B,\mathrm{true}}(t)|}{E_{B,\mathrm{true}}(t)+\epsilon}.
\end{equation}
Table~\ref{tab:ks_band_rollout} reports the mean of $\varepsilon_B(t)$ over the rollout window, and its value at the final time.
Two robust patterns emerge across trajectories: (i) low/mid bands remain relatively controlled, while (ii) dissipation-range bands exhibit huge \emph{relative} mismatch due to the surrogate maintaining a nonzero high-$k$ floor.

\begin{table*}[t]
	\centering
	\caption{Rollout band-energy error $\varepsilon_B(t)$ for three test trajectories for FNO. We report mean over time and final-time value. Bands: low $k\in[1,5]$, mid $k \in[6,15]$, hi1 $k \in [16,31]$, hi2 $k \in [32,47]$, hi3 $k \in [48,63]$, hi4 $k \in [64,95]$, hi5 k  $k \in [96,128]$.}
	\label{tab:ks_band_rollout}
	\begin{tabular}{llrrrrrr}
		\toprule
		Traj & metric & low & mid  & hi1  & hi2  & hi3  & hi4/hi5\\
		\midrule
		0 & mean & 2.03e-01 & 6.56e-02 & 3.80e-01 & 1.28e+00 & 8.27e+03 & 1.45e+08 / 8.58e+07\\
		0 & final& 4.06e-01 & 7.18e-01 & 4.48e+00 & 3.47e+00 & 3.27e+04 & 4.94e+08 / 1.81e+08\\
		\midrule
		1 & mean & 5.29e-01 & 1.02e-01 & 4.15e-01 & 1.23e+00 & 8.37e+03 & 1.23e+08 / 1.07e+08\\
		1 & final& 5.01e+00 & 3.72e-01 & 5.24e-01 & 6.33e-01 & 3.25e+03 & 3.12e+07 / 6.03e+07\\
		\midrule
		2 & mean & 2.51e-01 & 2.78e-01 & 1.07e+00 & 3.98e-01 & 2.48e+03 & 3.83e+07 / 1.90e+08\\
		2 & final& 7.38e-01 & 4.41e-01 & 1.18e-01 & 5.97e-01 & 7.12e+02 & 9.95e+06 / 2.32e+08\\
		\bottomrule
	\end{tabular}
\end{table*}

Figure~\ref{fig:ks} also shows the true/predicted/error heatmaps for a testing trajectory as well as the spatial profile $u(x)$ (true vs. predicted) and the spatial power spectrum.

A common explanation for the observed divergence  is that it is an inevitable consequence of chaos: specifically, that in a chaotic regime, pointwise trajectory comparisons are intrinsically ill-posed and therefore uninformative. While this caveat is important, our experiments indicate that it does not fully account for the discrepancies we observe. Across all tested architectures and training configurations, the growth rate of the prediction error consistently exceeded the rate implied by the system’s leading Lyapunov exponent, typically by a factor of a few. This systematic excess suggests that the dominant source of error amplification is not merely the intrinsic sensitivity of the underlying dynamics, but also model- and/or discretization-induced mechanisms that accelerate error growth beyond the baseline set by the true flow map’s local instability.

In addition to trajectory-level diagnostics, assessment of  predictive skill using a set of physically meaningful quantities of interest such as total and gradient energy ($\mathcal{E}(t)=\langle u(t,\cdot)^2\rangle_x$, $ \mathcal{G}(t)=\langle (\partial_x u(t,\cdot))^2\rangle_x$ further confirmed that spurious high-$k$ floor  manifests in statistics. Some of these effects can be mitigated by using Fourier Feature Networks  to modify the input encoding to boost high-frequency eigenvalues. This works for static function approximation but requires prior knowledge of relevant frequencies. For broadband, dynamically evolving spectra,  fixed encoding will not be appropriate.

\subsection*{Challenge problem: Combustors}

{The KS equation, while instructive, is a relatively mild multi-scale system. We now consider a far more demanding application to illustrate how spectral bias and coarse-graining interact in a setting of direct engineering relevance.}


Combustors for propulsion and energy generation represent  examples of a challenging multi-scale problem as they combine~\cite{Huang2023adaptive} compressible turbulence, chemical kinetics, and acoustics.
{From the standpoint of predictivity, combustors represent a worst case: the quantities of engineering interest (stability boundaries, mode transitions, reignition) depend on precisely the scales that spectral bias suppresses.}
Consider the example of Rotating detonation engines (RDE)~\cite{Raman2023RDE,Bennewitz2021Timescales} as an extreme case: the fastest relevant physics (chemical induction) occurs at $\tau_{\mathrm{chem}} \sim \mathcal{O}(10\text{--}100)\,\mathrm{ns}$, setting the Lyapunov timescale for detonation instabilities. The wave period is $\sim 250\,\mu\mathrm{s}$ and practical operating windows span $\mathcal{O}(1\text{--}10)\,\mathrm{ms}$. This yields $\sim 10^4$--$10^5$ Lyapunov times per operating interval, compared to $3$--$5$ for 10-day weather forecasts. The timescale ratio $\tau_{\mathrm{det}}/\tau_{\mathrm{chem}} \sim \mathcal{O}(10^3\text{--}10^4)$ implies that small-scale errors saturate almost instantaneously on the wave timescale. Unlike ERA5, where data assimilation filters the problematic small scales, the stiff chemistry \emph{cannot} be filtered without destroying the detonation physics. Neural surrogates trained on coarse observables (e.g., pressure traces) inherit none of the reaction-zone dynamics that govern mode transitions, re-ignition, and failure, precisely the phenomena of {\em predictive } engineering interest. If the goal is {\em reconstruction} of previously observed quasi-periodic flow features,   data-driven inference of linear and quadratic operators~\cite{Farcas2025Distributed} may be sufficient.

\textbf{Why Small Errors Become Large Failures.}
In combustors, high-$k$ fluctuations modulate mixing, flame wrinkling, and the effective time delay between heat release and acoustics.
Because thermoacoustic stability depends on phase, bias-induced delays of $10^{-4}$--$10^{-3}$ s can be qualitatively decisive when $T_{\text{ac}} \sim 10^{-3}$ s.
A model matching low-frequency pressure traces can still predict wrong growth rates, wrong saturation amplitudes, or wrong stability classifications: physically different outcomes from numerically similar training errors.

\section{Generative Models for Spectral Enhancement}

{Given the fundamental information loss identified in the previous section, a natural question is whether probabilistic methods can recover what deterministic surrogates cannot.}
A growing body of work integrates generative models with neural operators to address spectral deficiencies. Oommen et al.\ \cite{oommen2024diffusion} train diffusion models conditioned on neural operator outputs to sample high-frequency content, demonstrating improved energy spectra on Kolmogorov flow, high-Reynolds jet LES, and experimental Schlieren data. The approach stabilizes autoregressive rollouts by periodically correcting the spectral content that deterministic operators suppress. Jiang et al.\ \cite{Jiang2024DiAFNO} extend this to 3D turbulence with a combined neural operator-diffusion architecture. These methods achieve  improvements in \emph{distributional fidelity} and the generated fields exhibit  match target energy spectra across a broader wavenumber range.

However, a conceptual distinction must be maintained. Diffusion models sample from a learned conditional distribution $p(u_{\text{fine}} | u_{\text{coarse}})$; each sample is statistically consistent with the conditioning input but not uniquely determined by it. For applications requiring ensemble statistics or uncertainty quantification, this may be a good framework. For  trajectory prediction in chaotic systems, the generated fine scales might be statistically plausible but not dynamically coupled to the specific initial condition's history.

 Bhola and Duraisamy \cite{Bhola2024flowmatching} formulate flow matching in infinite-dimensional function space to learn a probabilistic transport from low-fidelity approximations to high-fidelity solutions via residual corrections. By parameterizing the flow vector field as a sum of linear and nonlinear operators and operating directly in function space, the approach achieves resolution-invariant inference while explicitly acknowledging the probabilistic nature of the correction. Generative models do not recover the ``true'' fine-scale field (which is non-unique given coarse observations), but offer a framework for principled sampling from a distribution of fields consistent with the available information. The distinction between learning a deterministic map and learning a conditional distribution is not a limitation to be overcome but a reflection of the underlying information-theoretic structure of the problem.

\section{Hybrid Methods: Resetting the Error Cascade}

{While generative models address the \emph{representational} gap, hybrid neural--classical methods tackle the \emph{accumulation} problem directly. The core idea is pragmatic: use the surrogate where it is fast and accurate (large-scale, short-horizon dynamics) and return to the full solver to regenerate the high-frequency content that spectral bias suppresses. This strategy offers gains in utility (net speedup) while bounding the predictivity loss that pure surrogate rollout incurs.}

Errors introduced by spectral bias or coarse-graining compound under autoregressive rollout until they corrupt even the well-resolved scales. A natural mitigation is to \emph{periodically reset} the surrogate by returning to a full-order solver, preventing error accumulation from reaching catastrophic levels.

Consider alternating between a neural surrogate $\mathcal{N}_\theta$ and a full-order solver $\mathcal{F}$: the surrogate advances $K$ steps before $M$ corrective steps with the full solver. The key insight is that the PDE solver does not merely correct errors, it \emph{regenerates the high-frequency content} that spectral bias systematically suppresses. This re-couples the resolved and unresolved scales, restoring dynamical consistency that pure surrogate rollout destroys. Instead of exponential error growth $\|\epsilon(t)\| \sim e^{\lambda_1 t}$ over the full trajectory, errors grow only over each surrogate window and are then quenched before the next window begins.

This strategy has origins in both reduced-order modeling and data assimilation. This has been demonstrated~\cite{Huang2023adaptive} in the context of projection-based ROMs for chaotic multiscale problems can maintain accuracy over long horizons by adaptively resampling the projection basis using short full-order solves when the ROM trajectory drifts outside the training manifold. The full-order samples serve a dual purpose: they correct the current state and enrich the reduced basis for future predictions. The same principle applies to neural surrogates—periodic full-order solves act as both error correction and implicit retraining signal, keeping the learned dynamics tethered to physical reality.

 In operational weather forecasting, models are periodically corrected by observations through analysis cycles. Techniques from the assimilation literature: nudging, ensemble spread monitoring, adjoint-based error estimation, can inform when to trigger corrections. A practical heuristic is to monitor the PDE residual $\|\mathcal{L}u_\theta - f\|$ or band-limited energy in dissipation-range wavenumbers; anomalous growth in either signals that the surrogate has drifted off-manifold and correction is needed.

Recent work explores related ideas. Kochkov et al.\ \cite{Kochkov2021} showed that learned corrections applied to coarse-grid simulations can match fine-grid accuracy for 2D turbulence, though corrections are applied at every step rather than interleaved.  List et al.\ \cite{List2024} use differentiable solvers to train neural networks that correct coarse-grid PDE solves, blurring the line between solver and surrogate. These approaches share the recognition that neural networks and classical solvers have complementary strengths: surrogates provide speed on smooth, large-scale dynamics while solvers provide the high-frequency regeneration and stability guarantees that surrogates lack.

The cost-benefit calculation is straightforward. If the surrogate costs $C_{\mathcal{N}}$ per step and the full solver costs $C_{\mathcal{F}} \gg C_{\mathcal{N}}$, then hybrid time-marching achieves net speedup whenever the surrogate window $K$ exceeds the correction window $M$ by a factor depending on the cost ratio. For typical claimed speedups of $C_{\mathcal{N}}/C_{\mathcal{F}} \sim 10^{-3}$, even modest surrogate windows yield substantial savings. The binding constraint is not cost but \emph{predictability}: $K$ must be chosen so that surrogate error remains acceptable, which depends on the Lyapunov time and spectral bias severity, not on training loss. A surrogate with excellent one-step error but severe spectral bias may support only small $K$, negating the cost advantage. Several open questions remain. Should switching be fixed-period or adaptive based on error indicators? Can the full solver correct only high-$k$ modes while the surrogate handles low-$k$ evolution, achieving scale-selective hybridization? Should surrogate parameters be updated online using full-order corrections as fresh training data? 

Other hybridization strategies are possible, such as replacing only the most expensive component of a multiphysics simulation with a neural surrogate. He et al.\ \cite{He2023icsheet} demonstrate this for ice-sheet dynamics: the Stokes equations governing ice velocity (the dominant computational cost) is replaced by a DeepONet, while the mass conservation equation for ice thickness evolution remains a standard finite element discretization (FEM). This physics-partitioned approach succeeds because the surrogate is asked to learn a \emph{static} operator (thickness and friction to velocity) rather than a dynamical map. The mass conservation equation, which controls the temporal evolution and where errors would compound, remains exactly satisfied by the FEM. The general question of under what conditions can hybrid schemes provably bound long-time error is open. In general,  defining a research program that takes neural surrogates seriously as \emph{components} of hybrid solvers rather than as wholesale replacements could be an active direction. {It has to be mentioned that hybrid approaches may require various levels of intrusive interaction with complex solvers (e.g.~\cite{Huang2023adaptive,Wentland25}) and introduce complex development challenges and will not be as portable and flexible as neural surrogates or other non-intrusive approaches (e.g.~\cite{Farcas2025Distributed}). This will be  an important consideration in practical settings.}

\section{Conclusions}

Neural surrogates for multiscale PDEs can be powerful if we separate
\emph{predictivity} (what is forecastable on the relevant horizon and QoIs) from
\emph{utility} (whether the method beats optimized classical baselines at fixed QoI tolerance,
including amortization).
	If the solution set lies on a low-dimensional, smooth manifold, many reduced models
	(neural or classical) will perform well. In many multi-scale problems, however,
	gradient-based training induces a low-pass learning dynamics (e.g., via NTK eigenvalue decay),
	so errors concentrate in the bands that control derivatives, dissipation, and phase.
	Small $L^2$ error can therefore coexist with large errors in physically decisive QoIs.
	A deterministic surrogate trained with MSE converges to a conditional mean, an over-smoothed field
	that may correspond to no true trajectory. What is identifiable is the \emph{conditional distribution}
	$p(u_{\mathrm{fine}} \mid u_{\mathrm{coarse}})$, motivating probabilistic (generative) corrections and
	uncertainty-aware evaluation rather than deterministic rollouts.

For chaotic systems, long-horizon trajectory prediction is typically ill-posed beyond a few Lyapunov times, and almost always, the error growth in neural surrogates is even more significant than Lyapunov growth.
Success is more likely when the target is (i) \textbf{statistics/invariant measures} or
(ii) \textbf{hybrid schemes} that periodically re-anchor the surrogate with a solver or observations to prevent
error cascades from corrupting resolved scales.

The strongest empirical critique of scientific ML reporting is that many papers compare against weak baselines, use mismatched hardware, or ignore amortization \cite{McGreivyHakim2024}.  For PDE surrogates, fair comparison  should be accompanied by: (1) problem characterization (intrinsic dimension/$n$-width proxy, scale separation,
and where possible a Lyapunov-time estimate); (2) test horizon in physical units (e.g., eddy turnovers/Lyapunov times);
(3) scale-aware metrics (band-limited errors, $H^1/H^2$ norms, and flux/dissipation-relevant quantities);
and (4) end-to-end cost curves versus optimized solver and ROM baselines with hardware parity.

{Despite the somewhat critical tone of this paper, the author acknowledges that many domains exist  where neural surrogates already offer value. Recognizing these domains is as important as understanding the limitations:}

\noindent \textbf{1. Smooth, low-dimensional parameter-to-QoI maps.}
Steady aerodynamics~\cite{Benedikt25} and design-space exploration where the solution manifold has small Kolmogorov width.
Here neural surrogates compete on amortized cost, not capability, but the cost advantage can be substantial for many-query applications. {(High utility; predictivity is straightforward because the manifold is benign.)}

\noindent \textbf{2. Learning invariant measures of chaotic systems.}
When the target is {\em stationary} statistics (PDFs, spectra, structure functions) rather than trajectories, neural operators~\cite{wang2024coarsegraining} can sample the attractor efficiently. It is noted that such capabilities have been demonstrated only in a limited set of examples. In systems with highly complex dynamics and limited availability of high-quality training data, current approaches may still face significant challenges. {(High utility when many-sample ensemble averages are needed.)}

\noindent \textbf{3. Subgrid modeling within resolved-scale frameworks.}
Neural networks as closure models \emph{within} a physics-based solver can accelerate coarse-grained simulations~\cite{Bhattacharya24}, provided the closure is not asked to reconstruct unresolved information, only to parameterize its statistical effect. {(Utility-driven: the solver retains predictivity while the surrogate reduces cost.)}

\noindent \textbf{4. Correction networks on residuals.}
Learning $\delta u = u_{\text{fine}} - u_{\text{coarse}}$ is often better-posed than learning $u$ directly, because the correction has smaller dynamic range and less scale separation~\cite{Bhola2024flowmatching}. More general classes multi-fidelity approaches will also be relevant here. {(Improves both predictivity and utility by reducing the approximation burden.)}

\noindent \textbf{5. Inverse problems and data assimilation.}
Here the ``blurring'' of deterministic surrogates can be acceptable because the goal is to find parameters or initial conditions consistent with observations, not to predict a unique trajectory. {(High utility; the relaxed predictivity requirement aligns well with surrogate strengths.)}

 The  path forward in more general tasks involving chaotic multi-scale dynamics may be to treat neural surrogates as components of hybrid, uncertainty-aware solvers, and certainly {\em not} as solver replacements.

\section{Acknowledgment}
This work was supported by the Center for Prediction, Reasoning, and Intelligence for Multiphysics Exploration (C-PRIME), a PSAAP-IV project funded by the Department of Energy, grant number DE-NA0004264 (program manager: David Etim).

\end{document}